# Nanomechanical Spectroscopy of 2D Materials

*Jan N. Kirchhof[1,*], Yuefeng Yu[1], Gabriel Antheaume[1], Georgy Gordeev[1,2], Denis Yagodkin[1], Peter Elliott[3], Daniel B. de Araújo[1], Sangeeta Sharma[3], Stephanie Reich[1] and Kirill I. Bolotin[1,*]*

[1] Department of Physics, Freie Universität Berlin, Arnimallee 14, 14195 Berlin, Germany

[2] Department of Physics and Materials Science, University of Luxembourg, 41 Rue du Brill, 4422 Belvaux, Luxembourg

[3] Max-Born Institute for Nonlinear Optics and Short Pulse Spectroscopy, Max-Born-Strasse 2A, 12489 Berlin Germany

*jan.kirchhof@fu-berlin.de    *kirill.bolotin@fu-berlin.de

**Abstract:**

**We introduce a nanomechanical platform for fast and sensitive measurements of the spectrally-resolved optical dielectric function of 2D materials. At the heart of our approach is a suspended 2D material integrated into a high Q silicon nitride nanomechanical resonator illuminated by a wavelength-tunable laser source. From the heating-related frequency shift of the resonator as well as its optical reflection measured as a function of photon energy, we obtain the real and imaginary parts of the dielectric function. Our measurements are unaffected by substrate-related screening and do not require any assumptions on the underling optical constants. This fast ($\tau_{rise} \sim 135\ ns$), sensitive (noise-equivalent power = $90\ \frac{pW}{\sqrt{Hz}}$), and broadband (1.2 – 3.1 eV, extendable to**

**UV – THz) method provides an attractive alternative to spectroscopic or ellipsometric characterization techniques.**

The interaction of light with a solid is encoded in the material's dielectric function $\epsilon_r(\omega) = \epsilon_1(\omega) + i\epsilon_2(\omega)$. Real and imaginary components $\epsilon_1, \epsilon_2$ contain the information regarding light absorption, propagation velocity, excitonic and plasmonic resonances, bandgaps, and many-body effects. The dielectric function is usually experimentally determined via spectroscopic ellipsometry,[1–4] a combination of reflection and transmission measurements,[5] or from spectrally-resolved reflection contrast (*dR/R*) using Kramers-Kronig relations.[6–8] Despite the broad applicability of these techniques, they are hard or impossible to apply in many situations. For example, optical measurements under oblique angles as required by spectroscopic ellipsometry are challenging at low temperature, ultra-high vacuum environments, and/or high magnetic fields. Measurements of transmission require large and thin samples on transparent substrates and may be affected by scattering. The Kramers-Kronig analysis requires broadband measurements of reflection and depends on empirical models of optical constants.[9]

For 2D materials, these problems become more severe. On one hand, 2D materials, in particular from the group of transition metal dichalcogenides (TMDs), feature a remarkable zoo of correlated phases including excitonic insulators,[10] Wigner crystals,[11,12] Bose Einstein condensates[13–15] and superconductors.[15] All these phenomena can be studied by analyzing the dielectric function. On the other hand, their observation requires uniform high-quality samples. Such samples are usually encapsulated in hexagonal boron nitride and have sizes in the micron range. Transmission or ellipsometry measurements of such nanostructures at ultralow temperatures or high magnetic fields are challenging.[16] Especially for studying plasmons or polaritons in patterned 2D materials in the form of photonic[17,18] or phononic crystals,[19] new optical characterization methods are needed. In addition, excitations in 2D materials are strongly screened by the underlying substrate. This screening perturbs the dielectric function also affecting the Kramers-Kronig analysis.

Here, we use nanomechanical spectroscopy[20–23] to accurately and quickly determine the optical dielectric function of 2D materials. For our proof-of principle experiments, we focus on few-layered

TMDs, well-understood materials with many pronounced features in their optical response. Our approach employs a suspended membrane made from the 2D material of interest. The mechanical resonance frequency of that membrane depends on its temperature, which, in turn, depends on the amount light absorbed by the material upon illumination. Therefore, by measuring changes of the resonance frequency of the membrane vs. the energy of photons ($E_\gamma = \hbar\omega$), we determine the absorption of the material. The membrane functions as its own photodetector – only sensitive to the amount of absorbed light and not, for example, to scattering and other losses. By combining the mechanically measured absorption with optically-recorded reflection, we restore the full dielectric function of the 2D material. We achieve very fast ($\tau_{\text{rise}} \sim 135\ ns$) and sensitive (noise-equivalent power = $90\ \frac{pW}{\sqrt{Hz}}$) measurements of the dielectric function for TMD materials in the range 1.2 – 3.1 eV. Our approach uses suspended samples – and therefore is unaffected by substrate-related screening and unwanted interaction with the probe beam. Furthermore, our approach does not require complex transmission measurements and therefore should function at low temperatures and high magnetic fields. Finally, by using the 2D materials itself for the detection of absorbed light, we can potentially obtain access to a broad spectral range from THz to UV allowing to study a large variety of materials.

Our first goal is to design a TMD-based resonator that controllably changes its frequency as it absorbs light. Such a resonator should have predictable mechanical resonances (the fundamental mode at frequency $f$) with small linewidth ($f_{\text{FWHM}}$) and thus high quality factor ($Q = \frac{f}{f_{\text{FWHM}}}$), linear mechanical response to illumination power, as well as high motion amplitude. To accomplish these goals, we design a hybrid resonator consisting of a TMD suspended over a circular hole in a high quality silicon nitride (SiN) membrane covered with a thin layer of gold. Our finite element method (FEM) simulations show, that the fundamental mode of such a resonator involves the TMD and suspended SiN oscillating together (Fig. 1a).[24,25] Absorption of light by the suspended TMD causes thermal expansion leading to a reduced tension $\sigma_0$ in the material and hence softening the resonance frequency of the entire system. Such a design presents several advantages:

First, the device features resonances with high quality factor at room temperature ($Q > 4000$). This is due to the mechanical quality and low losses in SiN and allows us to resolve frequency shifts with high resolution. This is contrast to 2D material-only resonators that show mechanical resonance frequencies with low quality factors ($Q \sim 100$).[26–29] Second, our design overcomes the disturbance typically associated with optical probing of mechanical resonances. To probe mechanical motion of our devices, we focus the probing beam on the suspended SiN area thereby avoiding heating of the material being measured. Third, in our system the mechanical resonances and their tuning are highly predictable, with device-to-device variations of ~5% or less. This allows us to simulate our systems with high confidence. The uniformity is also interesting from a technological point of view, as it can potentially allow coupling the oscillator to an external system, e.g. an LC-resonator for electrical signal amplification. Again, this is contrast to 2D material-only resonators that are affected by wrinkling and surface contaminations resulting in a large spread of mechanical resonance frequencies and unpredictable tuning.[29–31] Finally, by covering the SiN area with a thin layer of gold (also used for electrical actuation), we increase its reflectivity and thus enhance the signal to noise ratio. We note that the additional weight of the gold layer and high stiffness of SiN reduce the response of the hybrid mode to laser heating. We thus use micromechanical modelling to design a system giving a good compromise between signal strength and sufficient responsivity (Supplementary Fig. 2b,c), resulting in a device with a circular suspended TMD area of 10 µm diameter and a 20 µm square SiN window of 20 nm thickness. By using relatively low-stress SiN (240 MPa) we ensure that that the responsivity is large, and the sensitivity is improved.[23]

We realize the design described above by transferring a multilayer TMD (3 and 4 layers) onto a circular hole in a square SiN window covered with gold as shown in Fig. 1b (see Supporting Information Section 3). The mechanical motion of the hybrid SiN-TMD mode is excited by applying an AC+DC voltage between the device and a non-reflective gate ~40 µm below. The motion of the hybrid mode is detected interferometrically (Fig. 1c).[24] Critically, the probe beam (red beam in Fig. 1c) is focused onto the SiN area and does not significantly heat the device. A second laser source of tunable photon energy between 1.2 and 3.1 eV (blue beam in Fig. 1c) is focused on the suspended TMD area. The absorption of light from that beam increases the temperature of the device and downshifts the mechanical resonance, which we detect. All measurements are carried out in vacuum and at room temperature but are easily extendable to low temperatures and fiber-based setups. Fig. 1d shows the resonance response of the WSe$_2$ trilayer sample (device #1) without optical excitation. We find a prominent fundamental mode at $f_0 = 4.6702$ MHz with a $Q$~4500 and multiple higher order modes. We visualize the shape of the fundamental mode by scanning the probe laser across the device and plot the motion amplitude vs. position (Fig 1e, blue). The mode is well described by our simulations (Fig. 1a.) that give its shape (Fig. 1e, red) and frequency ($f_{0,sim}= 4.770$ MHz), confirming the well-controlled nature of our devices.

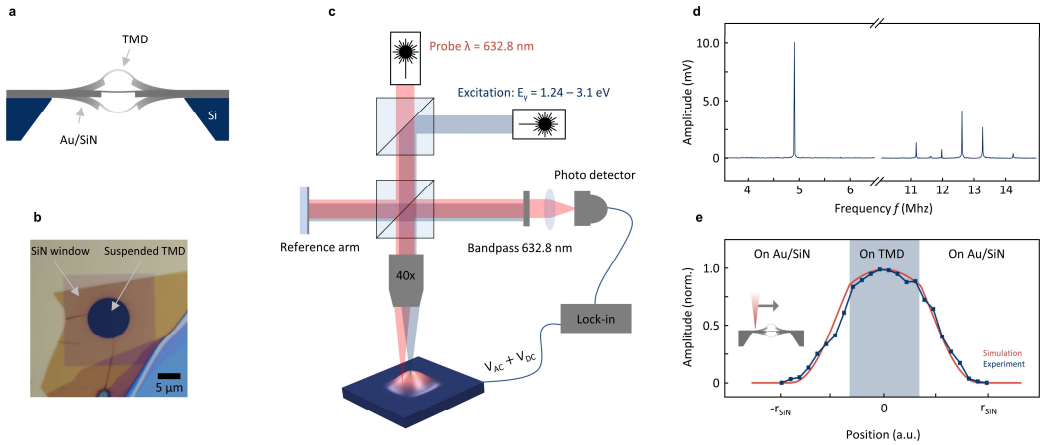

**Figure 1. SiN-TMD hybrid devices and interferometric motion detection. a**, Sketch of SiN-TMD hybrid resonance mode. The suspended SiN moves together with the TMD material and allows probing the resonances without focusing the probe laser on the TMD area. **b**, Optical micrograph of a WSe$_2$ trilayer flake suspended on a hole in a SiN window. **c**, Interferometric motion detection using a Michelson interferometer (red laser) with an additional broadly tunable laser (blue) to heat the TMD. The sample is placed in vacuum and the motion of the suspended area is actuated electrically. **d**, Measured amplitude vs. frequency for the trilayer WSe$_2$ SiN-hybrid device. The fundamental mode shows a large amplitude and enhanced quality factor. **e,** Relative motion amplitude of the fundamental mode extracted along a spatial line scan over the suspended area (blue) in comparison to simulation results (red). The amplitude of motion increases as the probe laser spot moves towards the center of the device and matches the simulated mode shape.

Our goal is to use our nanomechanical system to obtain a broadband absorption spectrum of the TMD. To this end we continuously vary the photon energy of the excitation laser and record the resonance response of the high $Q$ fundamental mode. In Fig. 3a we show the raw data of this measurement for the trilayer WSe$_2$ sample (device #1). Small laser powers ($P <$ 10 µW) are sufficient to cause clearly resolvable frequency shifts $\Delta f = f_0 - f$. The resonance frequency softens upon illumination, because the absorbed light heats the suspended TMD and reduces the built-in tension $\sigma_0$. From lower to higher energy, we observe an increasing down shift – in line with an overall increasing absorption towards the band gap of WSe$_2$. Within the spectra, there are multiple peaks and dips visible. Some of these features (e.g. the dip at 2.96 eV and the sudden jump at 1.91 eV) stem from variations in power of our excitation laser. We record the laser power at the sample position (Supplementary Fig. 5b) and use it to determine the resulting responsivity $\frac{\Delta f}{\Delta P}$ (Fig. 2b). As we show later, the measured responsivity is directly proportional to the optical absorption of the material and the distinct features associated with band-edge excitons of the TMD material are clearly noticeable. We identify the following excitonic transitions: A at 1.63 eV, B at 2.08 eV, A' at 2.28 eV, and B' at 2.68 eV. Positions and widths of theses peaks agree well with literature values.[6,32–34] As a reference, we perform photoluminescence measurements (PL) of the same device (Fig. 2b, grey). As expected, we find the peaks corresponding to A exciton and I exciton, with the latter corresponding to the indirect bandgap of multilayer TMD. For the A exciton, we find an expected small red shift and matching peak width, whereas the indirect peak is not visible in absorption due to the indirect nature of the transition.

Clear peaks are also apparent in the mechanical responsivity data for a second sample made from a different material from the TMD family (4L MoS$_2$ in Fig. 2c,d). Again, the positions and widths of these peaks match with excitonic transitions dominating optical absorption spectra of MoS$_2$.[35,36] We note that no assumption regarding material properties has been made in obtaining the spectra in Fig. 2b,d. In all measurements we observe a constant mechanical linewidth around the excitonic peaks (Supplementary Fig. 4a,b) and therefore exclude any cavity or material governed dynamic optomechanical back action effects.[37,38] We note that we can improve the measurement speed and sensitivity further by using a phase-locked loop (PLL) to directly measure the frequency shift vs. photon energy (Supplementary Fig. 5c). We obtain the same frequency response as shown in Fig. 2a, but now we record an entire spectrum in ~3 seconds, only limited by how fast the filter can change the output energy.

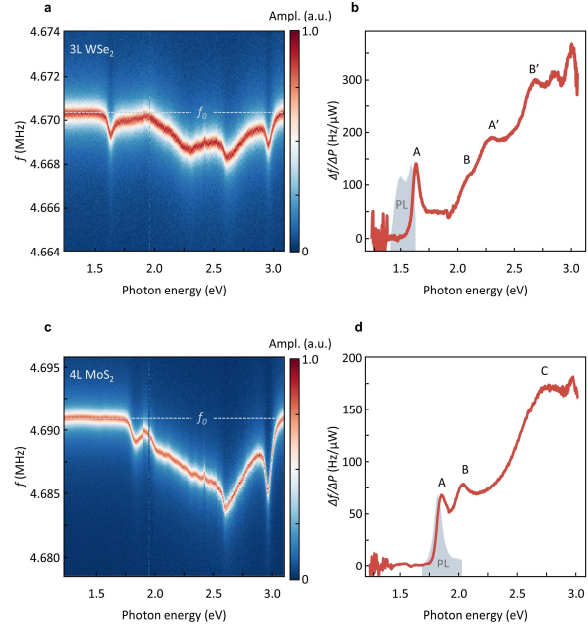

**Figure 2. Mechanical absorption spectroscopy in WSe$_2$ and MoS$_2$. a,c** Raw frequency response of the TMD-SiN-hybrid device as a function of photon energy for a WSe$_2$ and a MoS$_2$ samples. Multiple features are visible and towards higher energies the frequency shift increases as the absorption by the TMD increases. **b,d**, Responsivity $\frac{\Delta f}{\Delta P}$ vs. photon energy for WSe$_2$ and MoS$_2$. This signal is directly proportion to the absorption and shows clear excitonic features in both samples, which match reference PL measurements (grey). The MoS$_2$ sample was measured using higher laser power, causing larger absolute frequency shifts.

Next, we determine the optical dielectric function of the TMDs. For this, precise knowledge of both reflection $Refl(E_\gamma)$ and absorption $Abs(E_\gamma)$ in absolute units is required. To convert the measured frequency shift into absorption we model our hybrid system as two springs in series (elastic constants: $k_\text{SiN}$ and $k_\text{TMD}$) and link the responsivity (frequency shift $\Delta f$ normalized to laser power $\Delta P$) to the energy-depended absorption $Abs(E_\gamma)$ of the TMD via:

$$\Delta f \approx f_0 \frac{k_\text{TMD}}{2(k_\text{TMD}+k_\text{SiN})\sigma_\text{TMD}} \frac{\alpha E_{2D}}{1-\nu} \frac{\beta Abs(\lambda)}{h\kappa} \Delta P \qquad (1)$$

Here $m_\text{eff}$ is the effective mass of the mode, $\beta$ is a dimensionless factor determined by the temperature profile in the membrane, $\alpha$ is the thermal expansion coefficient, $E_\text{2D}$ is the 2D elastic modulus, $\kappa$ is the thermal conductivity and $\nu$ is the Poisson's ratio of the TMD (full derivation in Supporting Information Section 6). This expression shows that the responsivity is indeed proportional to the optical absorption. However, in our system, $m_\text{eff}$ and $\beta$ can only be obtained numerically and $k_\text{SiN}$, $k_\text{TMD}$ depend on a range of parameters (tension, thermal conductivity etc.) in a complex fashion. We thus use FEM modelling to determine the conversion factor linking $\frac{\Delta f}{\Delta P}(E_\gamma)$ to absorption. We first measure the frequency shift vs. incident laser power for 2.92 (425 nm) and 2.07 eV (600 nm) (Fig. 3a,b). For both energies we observe a linear and reversible downshift of the resonance frequency with increasing laser power (Fig. 3c). Having experimentally confirmed the linear behavior, we use the computed conversion factor (details on mechanical modelling in Supporting Information Section 1) to convert $\frac{\Delta f}{\Delta P}$ to absorption. For the fitted slopes from Fig. 3c (241±14 Hz/µW and 88±11 Hz/µW for 2.92 and 2.07 eV respectively), we get 30.4% and 11.2% absorption by the TMD membrane at respective energies. This is close to expectations.[6,33] As a benchmark for our approach we perform classical optical transmission measurements and find good agreement between the two methods of measuring absorption (see Supporting Information section 7). The simulation results also allow us to estimate the average temperature increase in the suspended TMD (Fig. 3c, right axis). For 30 µW laser power and 30.4% absorption, we find an increase to only a few degrees above room temperature, which is well inside the linear regime and below the TMD damage threshold.

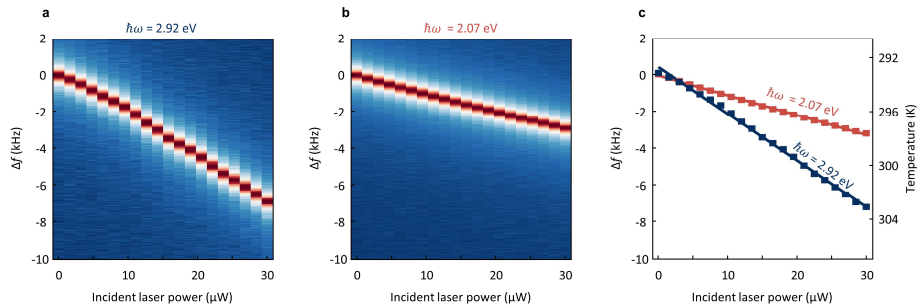

**Figure 3. Mechanical response to laser heating. a,b,** Frequency shift of the fundamental mode of a suspended trilayer WSe$_2$ SiN-hybrid device vs. incident laser power for 2.92 and 2.07 eV excitation. A fraction of light absorbed by the membrane causes heating, which reduces the built-in tension and softens the mechanical resonance. **c,** Fits to the frequency change vs. incident laser power, which by comparison to FEM-simulations allows to extract the absorption of the material at different energies. Right axis shows the average temperature in the suspended TMD obtained from such simulations.

Having extracted absorption, we are ready now to evaluate the dielectric function. In Fig. 4a,d we plot the spectrally resolved absorption (*Abs*) data for both samples obtained from the frequency shifts using the conversion factor (red traces). Next, we record reflection (*Refl*) off the TMD in the same sample configuration and same setup (blue traces in Fig. 4a,d, for measurement details see Supporting Information Section 8). We then use the transfer-matrix approach to relate reflected and absorbed light and obtain the complex refractive index $n$ and finally $\epsilon_1$ and $\epsilon_2$ by solving a system of equations at each measured energy (see Supporting Information Section 9):

$$\begin{cases} 1 - Abs - Refl = 1/|M_{11}(d, n(\epsilon_1, \epsilon_2))|^2 \\ Refl = |M_{21}(d, n(\epsilon_1, \epsilon_2))|^2/|M_{11}(d, n(\epsilon_1, \epsilon_2))|^2 \end{cases}$$

Here $d$ is TMD thickness and $M_{ij}$ are the matrix elements corresponding to the reflection and transmission from a thin suspended film (details in SI). No material parameters, except the thickness, are assumed in these formulas. The resulting $\epsilon_1$ and $\epsilon_2$ for 3L $WSe_2$ and 4L $MoS_2$ are plotted in Fig. 4b,c,e,f. Both real and imaginary parts of the dielectric function agree well with previous measurements.[1–4,6] Furthermore we find a close agreement with our ab-initio GW-Bethe Salpeter equation (GW-BSE) calculations (shaded in Fig.4b,c,e,f). The difference between measurements and theory for the real part of the dielectric function towards higher energies arises due to the known underestimation of the dielectric function by GW-BSE calculations in this regime. Using nanomechanical spectroscopy, we now have obtained full spectroscopic information of the material under test and thereby access to a majority of physical phenomena that make TMDs so exciting.

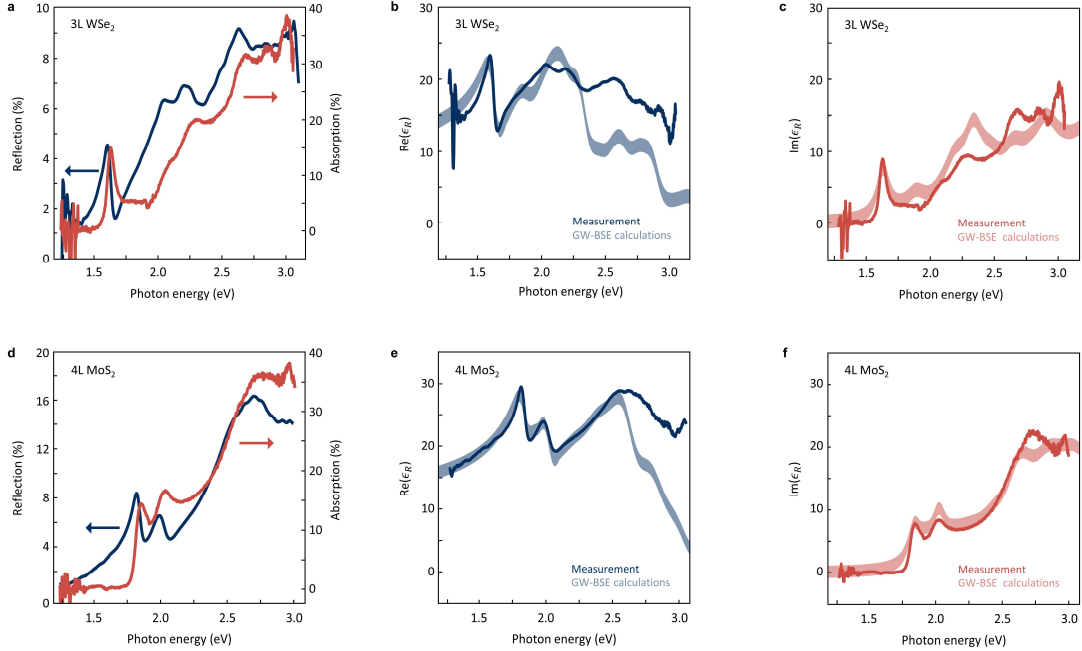

**Figure 4. Dielectric function of WSe$_2$ and MoS$_2$. a,d** Absorption (obtained from nanomechanical spectroscopty, red) and reflection (obtained optically, blue) for WSe$_2$ and MoS$_2$ vs. photon energy. Similar excitonic features are apparent in both measurements. **b,c,e,f** Real (blue) and imaginary (red) part of the dielectric function of both materials derived from absorption and reflection data (a,d). We find reasonable agreement for GW-BSE calculations (red, blue shaded).

How fast and how sensitive is our approach? The measurement speed is ultimately limited by the temperature equilibration time in the suspended TMD. To estimate this parameter, in Fig. 5a we plot the simulated average temperature of the suspended TMD vs. time after turning on the illumination (for a photon energy of 2.92 eV and 30 µW incident laser power). We extract a rise time of $\tau_{\text{rise}} \sim 135\ ns$, which corresponds to a bandwidth of 7.4 MHz, in line with experimental data for TMD devices of similar size.[39] To assess the sensitivity of our system, we determine the noise-equivalent power:

$$\eta = \frac{\sigma_A \sqrt{t} f_0}{\frac{\Delta f}{\Delta P}}, \qquad (2)$$

where $\frac{\Delta f}{\Delta P}$ is the responsivity (determined above), $\sigma_A$ is the Allan deviation obtained from time stability measurements (Fig. 5b,c) and $t$ is the sampling period.[40] For an optimal sampling period of 4 ms, we obtain $\eta = 90\ \frac{pW}{\sqrt{Hz}}$. This is only slightly higher than, for instance, state-of-the-art bolometers (2–100 $\frac{pW}{\sqrt{Hz}}$).[41–44] Nevertheless, the bandwidth is higher by several orders of magnitude for our 2D material-based system, which allows us to measure much faster.

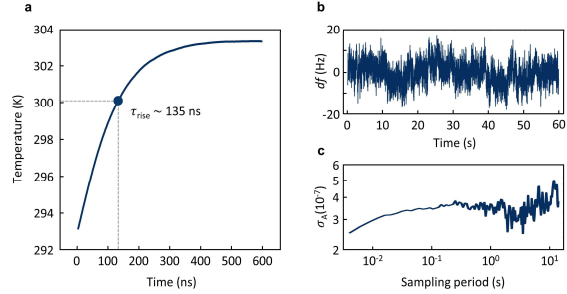

**Figure 5. Time resolution and sensitivity a,** Simulated time response of the average temperature in the suspended TMD as laser heating is introduced at t = 0 for 2.92 eV (30.4% absorption) and 30 µW laser power. We extract a rise time of 135 ns. **b,** Stability measurement of frequency vs. time without laser illumination, measured using a phase-locked loop. **c)** Extracted Allan deviation $\sigma_A$ vs. integration time in log-log scale derived from **b)**

Overall, our nanomechanical measurements are a new approach to obtain the dielectric function of 2D materials. This method is sensitive ($\eta = 90 \frac{pW}{\sqrt{Hz}}$), fast ($\tau_{\text{rise}} = 135\ ns$), and accurate. The absorption spectrum as well as the dielectric function for WSe$_2$ and MoS$_2$ are close to that obtained by others means. The approach does not require complex dielectric modelling or assumption about the material's optical properties (compared to e.g. constrained Kramer-Kronig analysis), optical transmission measurements (that are very difficult e.g. at low temperatures or high magnetic fields),[16] and works for samples the sizes of a diffraction-limited laser spot (unlike e.g. ellipsometry approaches). Moreover, our approach directly records optical absorption and is insensitive to scattering that is often hard to discriminate in all-optical techniques. By directly working with suspended samples, our approach is independent of tabulated optical constants of external materials and avoids perturbation of the excitonic features in 2D materials due to screening. Another advantage of using suspended samples is, that the incoming light is only absorbed by the materials under test. Finally, as the material serves as its own detector, it is relatively straightforward to extend the approach to different spectral ranges.

The sensitivity of our approach can be increased much further. First, the quality factor of optimized SiN membranes can reach $Q > 10^8$,[45] compared to around $10^3$ used here. This should likely result in a correspondingly higher measurement sensitivity. Second, Eq. 1 shows that the responsivity of our measurement is inversely proportional to the thermal conductivity of the 2D material. This quantity can be reduced by e.g. patterning the membrane into a trampoline shape[44] or e.g. via defect-engineering.[46,47] Third, at low temperatures the mechanical quality factor increases while the ratio of thermal conductivity and thermal expansion entering Eq.1 stays roughly constant. Therefore, we expect the low temperature sensitivity of our approach to increase. As most 2D materials show comparable mechanical and thermal properties, we expect our approach to universally work well. Graphene and hBN, however, will show frequency shifts in the opposite direction, as they have a negative thermal expansion coefficient. For them we also expect slightly reduced sensitivity due to their higher thermal conductivity. The combination of 2D materials with low loss SiN results in enhanced Qs and therefore provides an attractive platform for studying optomechanical cooling and self-oscillation phenomena – especially at low temperatures. Finally, we hope to implement our measurement scheme completely on a chip using electrical readout to create a compact, robust, and highly sensitive nanomechanical platform for spectroscopic characterization of 2D materials.

## Supporting Information

Details of the FEM-simulations, AFM force indentation measurements, overview of measured samples, details on interferometric motion detection, discussion on dynamical back-action effects, springs in parallel model (derivation of Eq.1), optical transmission measurements, details on reflection measurements, obtaining the dielectric function for thin films, RPA and BSE calculations, measuring the Allan deviation, calculating the sensitivity and details on the photoluminescence measurements.

## Acknowledgements

This work was supported by Deutsche Forschungsgemeinschaft (DFG, German Research Foundation, project-ID 449506295 and 328545488), CRC/TRR 227 (project B08 and A04), ERC Starting grant no. 639739 and CSC 202006150013.

# Supplementary Information: Nanomechanical Spectroscopy of 2D Materials

*Jan N. Kirchhof[1*], Yuefeng Yu[1], Gabriel Antheaume[1], Georgy Gordeev[1,2], Denis Yagodkin[1], Peter Elliott[3], Daniel B. de Araújo[1], Sangeeta Sharma[3], Stephanie Reich[1] and Kirill I. Bolotin[1*]*

[1] Department of Physics, Freie Universität Berlin, Arnimallee 14, 14195 Berlin, Germany

[2] Department of Physics and Materials Science, University of Luxembourg, 41 Rue du Brill, 4422 Belvaux, Luxembourg

[3] Max-Born Institute for Nonlinear Optics and Short Pulse Spectroscopy, Max-Born-Strasse 2A, 12489 Berlin Germany

*jan.kirchhof@fu-berlin.de    *kirill.bolotin@fu-berlin.de

1. **Finite element method (FEM) simulations**

For the FEM modelling we use the structural mechanics module of COMSOL Multiphysics Version 5.5. We build a model around the suspended area partially including the silicon support (Fig. S1a) and use a swept mesh for the thin layers with high density around for TMD flake and SiN window (Fig. S1b). To determine the resonance frequency and mode shape (Fig. S1c) we conduct a prestressed eigenfrequency study. To include the effect of laser heating, we add a study step to implement a Gaussian heat source (Fig. S1d, 30.4% absorption and 30 µW laser power – comp. Fig. 3a main paper) and calculate the heat profile upon laser heating of the center of the suspended TMD (Fig. S1e). This allows us to determine the conversion factor, which captures the tuning of the fundamental mode with laser power, following Eq. 1 from the main paper. The conversion factor slightly depends on wavelength of the heating laser because the laser spot size varies with wavelength, what results in a slightly different heat profile in the suspended TMD. To account for this, we measure the spot size of heating laser at different wavelengths and use this as input for our simulations. In Fig. S1f we plot the conversion factor for device #1 (3L $WSe_2$). For device #2 (4L $MoS_2$), we obtain a conversion factor in the range of 461 to 476 Hz/µW showing comparable scaling with wavelength as device #1. The difference between devices here is due to different thermal conductivities, hole sizes and layer thickness between devices.

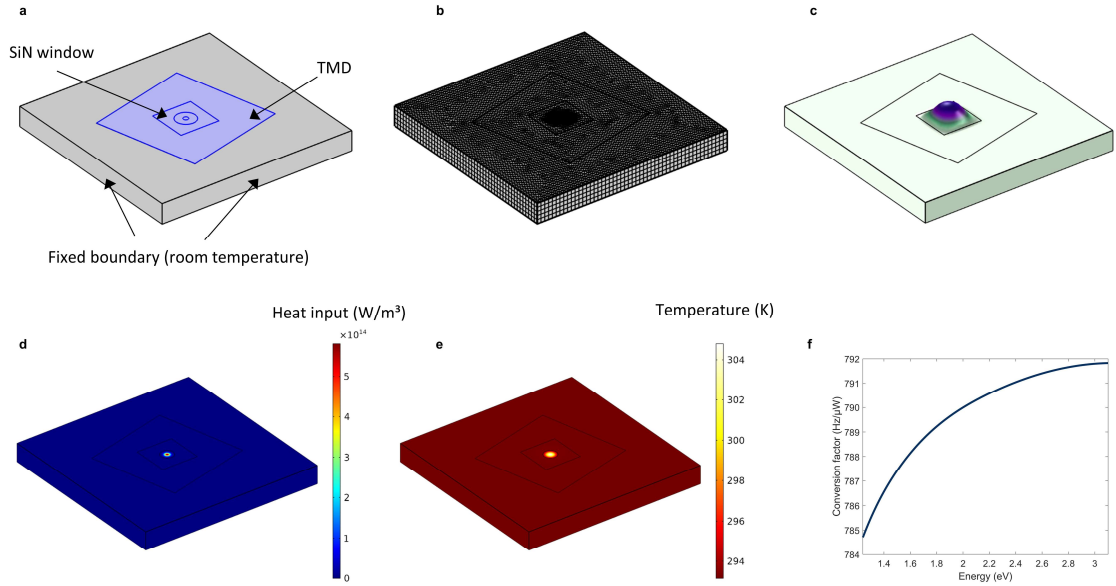

**Figure S1 FEM simulations of SiN-TMD hybrid devices a)** Model geometry with a thin layer of TMD (blue) placed on the SiN window **b)** Corresponding mesh, all thin layers are meshed as swept layers with high density **c)** Simulated fundamental mode of the hybrid system **d)** Heat input in the shape of a gaussian beam to simulate the effect of laser heating **e)** Resulting heat profile upon 30 µW incident laser power and 30.4 % absorption (corresponds to a photon energy of 2.92 eV for the 3L WSe$_2$ sample) **f)** Obtained conversion factor for Device #1 (3L WSe$_2$).

In order to optimize the dimensions of the SiN window, we simulate the driven mechanical resonances in the frequency domain. We start by simulating a circular TMD-only drum resonator (diameter 10 µm) as reference and adjust the isotropic damping to match the experimental Q for such resonators (~100). We then simulate the entire hybrid device (including SiN and gold). In Fig. S2a we plot the simulated displacement vs. frequency for the hybrid device probed on the SiN area, 2 µm away from the suspended TMD area. Again, we adjust the isotropic damping in SiN and gold to match experimental values. We now vary the SiN window size and extract the amplitude of motion at constant drive (signal strength, plotted in Fig. S2b). As expected, larger devices oscillate at large amplitudes providing more signal. Nevertheless, while oscillating at a higher amplitude, larger devices are less responsive to heating. Indeed, in Fig. S2c we plot the relative responsivity (change of resonance frequency for a constant laser heating) vs. window size. Combining the insights from Fig. S2b,c we choose a window size of 20 µm as a reasonable compromise between high vibrational amplitude and high responsivity.

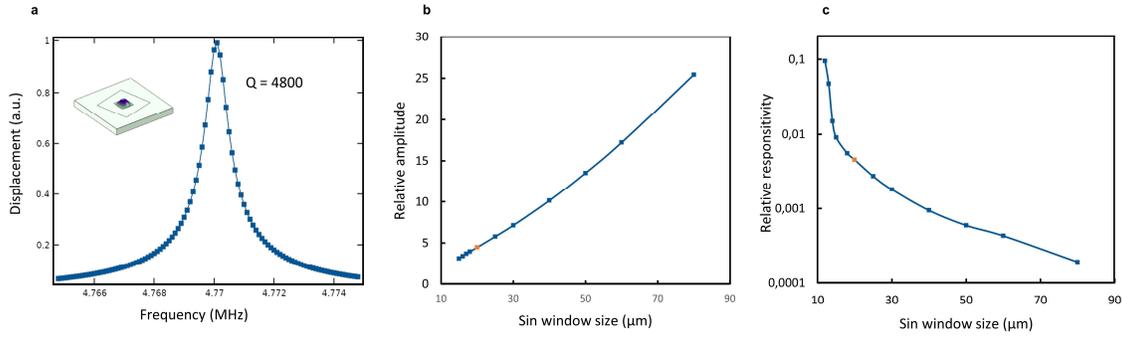

**Figure S2 Finding ideal device parameters a)** Simulated mechanical motion of the hydride with Q matching experimental results **b)** Simulated amplitude (signal strength) vs. SiN window size **b)** Relative responsivity to laser heating vs. SiN window size. We choose a window size of 20 µm (orange spot) as a compromise between high responsivity and sufficient amplitude amplification

All material properties used in our simulations are summarized in table 1. For quantities that show a large spread in the literature values (values for the TMD materials in particular) we used average values. In general, we preferably choose experimental references for suspended samples of the suitable layer thickness.

| Material | Quantity | Value | Reference |
|---|---|---|---|
| $MoS_2$ | Young's modulus $E$ | 330 GPa | [1] In agreement with AFM force-indentation measurements (see below) |
| | Poisson's ratio $\nu$ | 0.125 | [1] |
| | Density $\rho$ | 5060 kg/m³ | [2] |
| | Thermal conductivity $\kappa$ | 60.3 W/(m·K) | [3–6] |
| | Thermal expansion coefficient $\alpha$ | $7.6 \cdot 10^{-6}$ 1/K | [7] [8] |
| | Heat capacity at constant pressure $c_p$ | 397 J/(kg·K)) | [9,10] |
| | Built-in stress (tension) $\sigma_0$ | 44.7 MPa ($\sigma_{2D} = 0.11$ N/m) | Force-indetantion AFM |
| | Layer thickness $d$ | 0.615 nm | [11] |
| WSe$_2$ | Young's modulus $E$ | 167.3 GPa | [12] In agreement with AFM force-indentation measurements (see below) |
| | Poisson's ratio $\nu$ | 0.19 | [13] |
| | Density $\rho$ | 9320 kg/m³ | [14] |
| | Thermal conductivity $\kappa$ | 26.5 W/(m·K) | [15] |
| | Thermal expansion coefficient $\alpha$ | $7 \cdot 10^{-6}$ 1/K | [7] |
| | Heat capacity at constant pressure $c_p$ | 188 J/(kg·K) | [9] |
| | Built-in stress (tension) $\sigma_0$ | 46.2 MPa ($\sigma_{2D} = 0.09$ N/m) | AFM force indentation |
| | Layer thickness $d$ | 0.651 nm | [11] |
| Au | Young's modulus $E$ | 78.5 GPa | [16] |
| | Poisson's ratio $\nu$ | 0.42 | [16] |
| | Density $\rho$ | 19300 kg/m³ | [16] |
| | Thermal conductivity $\kappa$ | 312 W/(m·K) | [16] |
| | Thermal expansion coefficient $\alpha$ | $14 \cdot 10^{-6}$ 1/K | [16] |
| | Heat capacity at constant pressure $c_p$ | 130 J/(kg·K) | [16] |
| | Built-in stress | 160 MPa | [17] |
| SiN | Young's modulus $E$ | 232 GPa | [18] |
| | Poisson's ratio $\nu$ | 0.23 | [18] |
| | Density $\rho$ | 2810 kg/m³ | [19] |
| | Thermal conductivity $\kappa$ | 31 W/(m·K) | [18] |
| | Thermal expansion coefficient $\alpha$ | $2.55 \cdot 10^{-6}$ 1/K | [18] |
| | Heat capacity at constant pressure $c_p$ | 887 J/(kg·K) | [18] |
| | Built-in stress | 240 MPa | Norcada (manufacturer) |

| Si | Young's modulus $E$ | 160 GPa | 20 |
|---|---|---|---|
| | Poisson's ratio $\nu$ | 0.27 | 20 |
| | Density $\rho$ | 2330 kg/m³ | 20 |
| | Thermal conductivity $\kappa$ | 160 W/(m·K) | 20 |
| | Thermal expansion coefficient $\alpha$ | $3 \cdot 10^{-6}$ 1/K | 20 |
| | Heat capacity at constant pressure $c_p$ | 692 J/(kg·K) | 20 |
| | Pretension | 0 | Irrelevant for simulations |

## 2. AFM force indentation

One crucial parameter, which is known to vary from device to device is the built-in tension and 2D elastic modulus. To eliminate this uncertainty in our simulations, we perform force indentation measurements in the centre of the membrane (following Ref. [21]) and extract the built-in tension and 2D elastic modulus for each sample. We use cantilevers of intermediate stiffness (k~3 N/m) and only apply small loads (150 nN) to avoid damaging the sample. In Fig. S3a we show a force-displacement-curve for device #1. We account for cantilever bending and deformation of the SiN membrane. We fit a curve following:

$$F = \pi \sigma_{2D} d + q^3 \frac{E_{2D}}{a^2} d^3 \qquad (S1)$$

Here $a$ is the radius of the drum, and $q = \frac{1}{1.05 - 0.15\nu - 0.16\nu^2}$ is a dimensional factor dependent on the Poisson's ratio $\nu$ ($q = 0.98$ for WSe$_2$ and q = 0.97 for MoS$_2$). For a range of samples, we find a linear dependence on tension with layer thickness (Fig. S3b). We attribute the observed homogeneity to the cleanliness and uniformity in our samples after annealing (comp. Fig. S4).

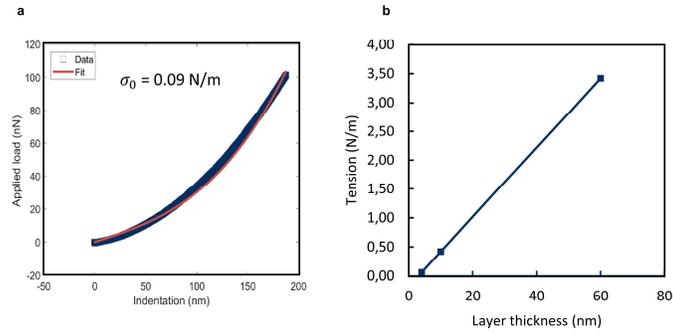

**Figure S3 AFM force indentation to determine pre-tension. a)** Force vs. displacement as well as a fit to Eq. S1. We extract a pretension of roughly 0.1 N/m for most our devices **b)** Statistics on pre-tension vs. thickness. We find a linear relation between pre-tension and layer thickness in our devices.

## 3. Sample fabrication and overview

To fabricate our hybrid devices, we transfer TMDs onto a circular hole using the all-dry PDMS method.[22] The SiN chip is beforehand covered with a thin layer of gold (30 nm) to electrically contact the TMD and to increase its reflectivity. After transfer we perform an annealing step (3 h, 200 °C) in vacuum to remove residues and assure good adhesion to the substrate. We fabricate and measure multiple samples. Microscope images and AFM topography scans for device #1-3 are shown in Fig. S4a,b,d,e,g,h. For all samples (Fig. S4c,f,i), we find a high Q fundamental mode of almost constant (except device #3, which has a thicker gold layer). There are some variations in frequency, because the hole size and gold thickness are different for the devices. Our simulations (grey dashed lines) describe the measured frequencies well (Fig. S4c,f,i).

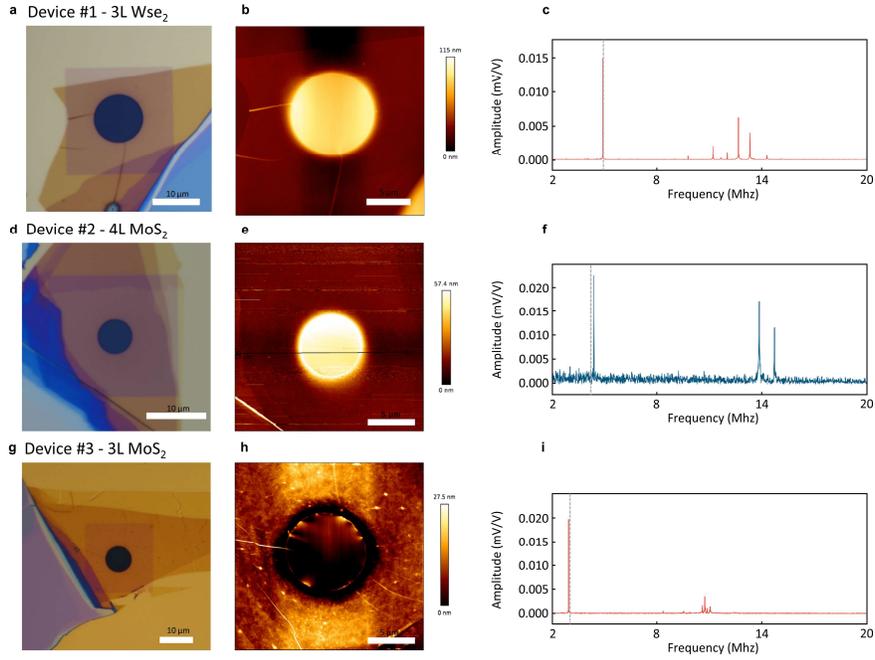

**Figure S4 Sample overview.** Microscope images **(a,d,g)** and AFM topography **(b,e,h)** of device #1-3. The samples are uniform and well attached to the substrate **c,f,i)** Displacement (amplitude) vs. frequency for device #1-3. We find a dominant high-Q fundamental mode for all samples. The resonance frequencies match with simulated values (grey dashed lines).

## 4. Details on interferometric motion detection

The sample is placed upside down in a vacuum chamber of < $10^{-5}$ mbar. By applying a DC+AC voltage relative to the non-reflective gate electrode, we mechanically actuate the suspended area of the chip. The motion of the TMD is detected by a Michelson interferometer. We focus a 632.8 nm HeNe laser (<1 µW, ~1.5 µm spot size) on the SiN area of the sample and the reflected light is superimposed with the light coming from the reference arm and guided into an avalanche photodetector. The resulting interference signal is highly sensitive to relative displacements and allows us to detect the motion of suspended samples. We actively stabilize the relative position of the reference arm via a mirror on a piezo and thereby ensure constant interference conditions and good signal strength over a large period of time. In addition to the probe laser, we implement an excitation laser of tunable colour (1.2 – 3.1 eV, blue in Fig. 1c). We use a band pass filter (BP) to block the excitation laser from reaching the detector and overloading it. The large separation (40 µm) of the non-reflective gate and sample negates all cavity-related optomechanical backaction effects and allows us to measure purely static heating effects in our sample over a very large range of photon energies.

The interferometric setup is shown in detail in Fig. S5a. Along the beam path of the probe laser, we first implement an optical isolator to avoid back reflected light into the laser, which can cause instabilities and power fluctuations. The beam is then expanded to completely fill the objective (40x 0.6NA). In a first beam splitter we add light from the excitation laser and in a second beam splitter, we guide half the light towards the reference arm and half through the objective onto the sample in a vacuum chamber. The relative position of the reference arm to the sample determines the amplitude of the interferometric signal. We use a piezo electric element to control this distance and stabilize the system using a PID-loop locked to a small reference signal at 941 Hz sourced by Lock-In amplifier (Zurich Instruments MLFI). The sample in the vacuum chamber is clamped upside down onto our sample holder and with a spacing of roughly 40 µm, we place our grounded gate electrode. Electrical driving is realized by mixing a DC voltage (210 V, supplied by a Keithley source meter) with an AC component (typically -5 dBm) from our vector network analyzer (VNA, Agilent E5071C) in a high voltage Bias T (Particulars BT-01) and applying it to the gold layer of the sample, which contacts the TMD flake. For smaller frequency ranges and phase-locked-loop (PLL) measurement, we use a lock-in amplifier (Zurich Instruments MFLI). In Fig. S5b we show the power spectra of our excitation laser source (measured at the sample position) with different neutral density filters (ND) implemented, which are used to calculate the relative frequency shifts $\frac{\Delta f}{\Delta P}$. We perform a small linear correction (order of Hz) to account for temperature changes in the room during measurements of the maps (Fig. 2 a,c). In the PLL-configuration (25 kHz bandwidth) we can measure the heating induced frequency shifts $\Delta f$ quickly and with high sensitivity even at low laser powers (raw data for ND 1.5 in Fig. S5c).

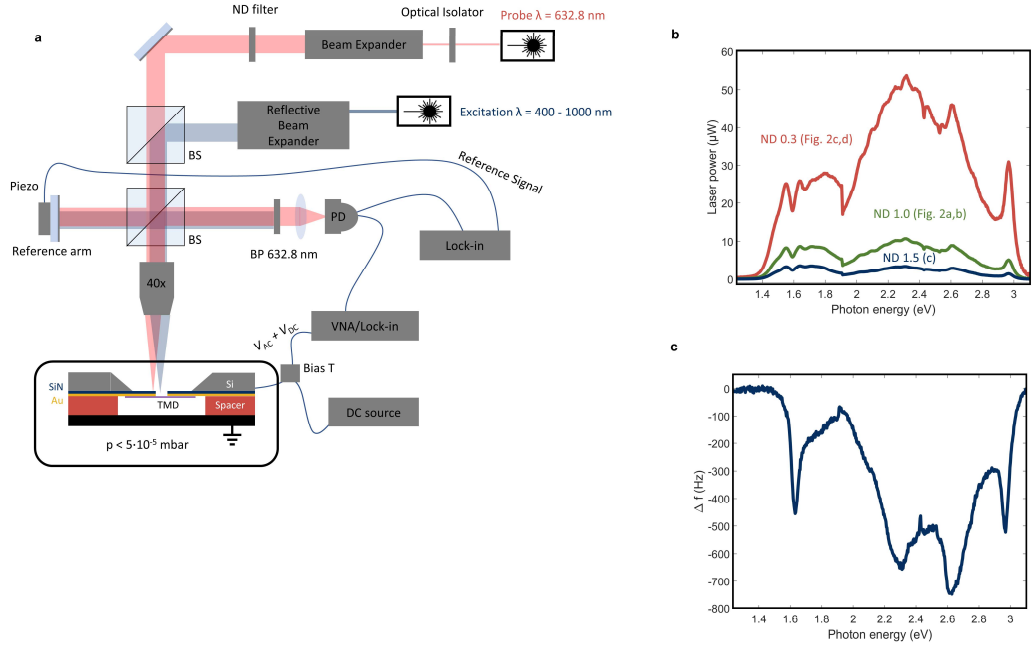

**Figure S5 Setup details and PLL-data a)** Detailed sketch of setup and **b)** Measured output of the tunable excitation laser at the sample position vs. wavelength. This data is used to normalize the frequency shift. **c)** PLL measurements of device #1, using the laser power plotted in **b)**.

## 5. Consideration of dynamical back-action effects

In nanomechanical resonators also dynamic optomechanical back-action (in contrast to static heating) effects can alter the resonance frequency ($f$) and its FWHM ($f_{FWHM}$) especially at large laser powers.[23–25] This occurs e.g. in cavity interferometers, where the laser power, which the oscillating membrane is exposed to, varies significantly over a short spatial distance.[23,24] For this a reflective surface close to the moving membrane is needed.[23,24] The effects furthermore only occur when the spatial symmetry is broken due to deforming the membrane out of plane.[23,24] In our system the gate is non-reflective and far away from the membrane (~ 40 µm). Additionally, the applied electrostatic pressure by the gate voltage is relatively small and SiN-TMD hybrid system rather stiff, so there is no breaking of symmetry in out of plane direction. Considering all the points above, we can exclude cavity related back-action effects in our system.

Also, strain-induced shifts in absorption in the material itself can cause dynamic back-action effects.[25] Here again a breaking of symmetry, large laser powers and soft systems (small spring constant) are needed. We therefore also exclude material related back-action effects.

To verify this experimentally we extract $f_{FWHM}$, whilst illuminating the sample at different wavelengths (Fig. S6a,b). If there were any dynamic back-action effects influencing the system, the $f_{FWHM}$ should show significant variations.[23–25] We do not observe such variations and thereby experimentally confirm the absence of dynamic back-action effects.

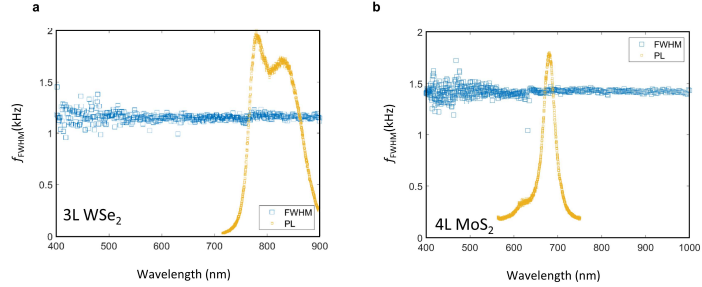

**Figure S6 Reference measurements check for dynamic back-action effects. a,b)** FWHM vs. wavelength for device #1 (WSe$_2$) and #2 (MoS$_2$) and photoluminescence measurements as reference for the excitonic resonances. We observe a constant FWHM over the entire wavelength range and thereby experimentally exclude dynamic optomechanical back action effects.

### 6. Springs in parallel model (derivation of Eq.1)

The goal of equation 1 from the main text is to intuitively relate the heating-induced change of the overall resonance frequency of our resonator to tension/frequency changes of its components, i.e. the TMD alone and the SiN alone. That expression is important for developing a qualitative understanding of our system, whilst we use FEM simulations to capture the complex device geometry for all quantitative evaluations and results shown in the main text. We express frequencies (*f*) and frequency changes ($\Delta f$) of each resonator upon illumination/heating via effective spring constants defined as $k = 4\pi^2 f^2 m_{eff}$, where $m_{eff}$ is the numerically determined effective mass[4] (see Fig. S7). So, the question we would like to answer: is there a simple expression relating effective spring constants of the SiN ($k_{SiN}$), the TMD ($k_{SiN}$), and the compound system ($k_{total}$)?

As a first step towards obtaining a simple model, we numerically obtain the frequency and frequency changes upon illumination (power 30 µW; 30.4 % absorption) of our 3 resonators using detailed FEM-simulations of the experimental geometry (see Fig. S7 a-c). From the resonance frequencies and effective masses, we find: $k_{TMD} \approx$ 0.52 N/m, $k_{SiN} \approx$ 50.43 N/m and $k_{total}$ =65.48 N/m. For the heating-induced frequency changes upon laser illumination with 30 µW laser power and 30.4% absorption (corresponds to the measurement in Fig. 3a @ 2.92 eV), we find: $\Delta k_{TMD} \approx -0.16 \frac{N}{m}$, $\Delta k_{SiN} \approx -0.01 \frac{N}{m}$ and $\Delta k_{total} \approx -0.20 \frac{N}{m}$. These simulations match experiments: from the experimentally measured heating-induced frequency shift, we extract:

$$\Delta k_{total}^{Exp} = -k_{total}^{Exp}\left(1 - \frac{(f-\Delta f)^2}{f_0^2}\right) = -64.45 \frac{N}{m}(1 - 0.9970) \approx -0.19 \text{ N/m}$$

(with $f_0 = 4.67$ MHz; $\Delta f = 7.2$ kHz). Moreover, $\Delta k_{total}$ linearly depends on $\Delta k_{TMD}$. We see that our numerical results can be described with reasonable precision by a simple formula, $k_{total}=k_{TMD}+k_{SiN}$. This formula corresponds to effective springs of the TMD and the SiN connected "in parallel". In fact, this expression can be derived analytically in a 1D toy model of the combined TMD/SiN resonator (see Fig. S7d). We start by assuming that the hole in the SiN does not affect the effective elastic constants of SiN (good approximation when the hole is small) and simplify the membrane profile for ease of estimates (Fig. S7d). In this geometry we can approximate the extension of the central point of the membrane as:

$$\delta x \approx \frac{x^2}{2L} \quad \quad (S2)$$

Approximating that both TMD and SiN have the same strain ($\epsilon$), we can now define the potential and kinetic energy as following:

$$E_{pot} = 2\sigma\delta x = 2\epsilon(h_1 E_{2D}^{SiN} + h_2 E_{2D}^{TMD})\delta x = \frac{\epsilon}{L}(h_1 E_{2D}^{SiN} + h_2 E_{2D}^{TMD})x^2 \quad (S3)$$

$$E_{kin} = \frac{m(\delta \dot{x})^2}{2} \approx \frac{m(\dot{x})^2}{2} \quad (S4)$$

From the conservation of energy, we obtain the angular frequency of the harmonic motion of the combined system:

$$\omega^2 = \frac{2\epsilon(\frac{\epsilon}{L}(h_1 E_{2D}^{SiN} + h_2 E_{2D}^{TMD}))}{L m_{eff}} \quad (S7)$$

Defining an effective spring constants for the resonator, with $k_{SiN/TMD} = \frac{2\epsilon}{L}(h_{1/2} E_{2D}^{SiN/TMD})$, we see that the resonance frequency of the combined system can be expressed as:

$$f_0 = \frac{1}{2\pi\sqrt{m_{eff}}}\sqrt{k_{SiN} + k_{TMD}} \quad (S8)$$

This is exactly the resonance frequency of the resonator with TMD and SiN springs "in parallel".

Next, we will look at changes in frequency ($\Delta f = f(T) - f_0$) caused by laser heating. With the heating laser turned on, light is absorbed, and the TMD resonator heats up and overall softens. The SiN is well heat sunk via the gold layer. Our simulations show that its temperature and stress remain almost constant ($\Delta T \approx 0.03$ K, $\frac{\Delta \sigma}{\sigma_0} \approx 0.02\%$, for 30.4% absorption and 30 µW incident laser power). Therefore, we assume that $k_{SiN}$ is temperature-independent and express the resonances frequency with laser heating as:

$$\Delta f = \frac{1}{2\pi\sqrt{m_{eff}}}\sqrt{k_{SiN} + k_{TMD} - \Delta k_{TMD}} - f_0 \quad (S9)$$

We now expand the term to first order for $\frac{\Delta k_{TMD}}{k_{TMD}}\bigg|_{SiN} \ll 1$ and obtain

$$\Delta f = f_0 \left(\sqrt{1 - \frac{\Delta k_{TMD}}{k_{TMD} + k_{SiN}}} - 1\right) \approx f_0 \frac{\Delta k_{TMD}}{2(k_{TMD} + k_{SiN})} \quad (S10)$$

For the "TMD-spring", we can relate the change in spring constant to a change in built-in tension:

$$\frac{\Delta k_{TMD}}{k_{TMD}} = \frac{\Delta \sigma}{\sigma_0} \quad (S11)$$

The change in tension due to thermal expansion is given by:

$$\Delta \sigma = \frac{\alpha E_{2D}}{1-\nu}\Delta T, \quad (S12)$$

Where $\alpha$ is the thermal expansion coefficient, $E_{2D}$ is the 2D elastic modulus and $\nu$ is the Poisson's ratio of the TMD. The change in temperature $\Delta T$ is proportional to the amount of absorbed laser power:

$$\Delta T = \frac{\beta Abs(\lambda)}{h\kappa}\Delta P, \quad\quad (S13)$$

where $\kappa$ is the thermal conductivity, $h$ is the thickness of the membrane and $\beta$ is a pre-factor determined by the temperature profile in the membrane. Combining Eq. S11-13 we obtain Eq. 1 from the main text:

$$\Delta f \approx f_0 \frac{k_{TMD}}{2(k_{TMD}+k_{SiN})\sigma_{TMD}} \frac{\alpha E_{2D}}{1-\nu} \frac{\beta Abs(\lambda)}{h\kappa} \Delta P \quad\quad (S14)$$

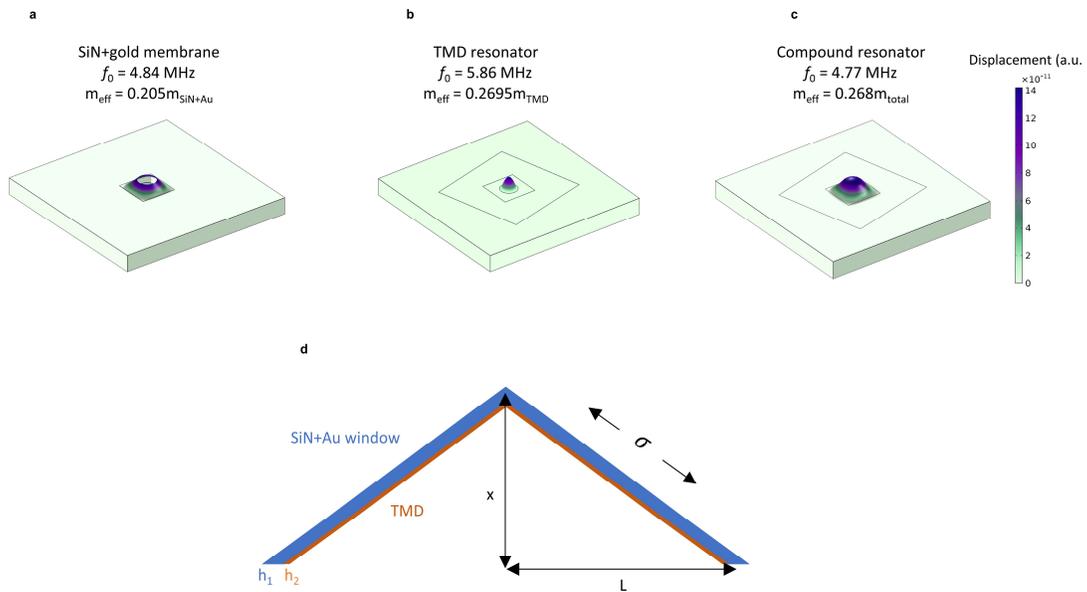

**Figure S7 Numerical evaluation of spring constants and springs in parallel model. a-c)** Mode shape of the individual resonances of the substrate (SiN+gold), the TMD resonator and the combined system. **d)** Sketch for simplified model describing the springs in parallel.

## 7. Transmission measurement as benchmark

In order to validate our nanomechanical measurement approach and backup our simulations, we perform optical transmission measurements with an objective below and above the sample (Fig. S8a).

We use a broadband white light laser and measure the transmission through the sample (Fig. S8b), deduct the dark counts and normalize to an empty hole without any TMD material to obtain the transmission:

$$T = \frac{T_{sample}-T_{dark}}{T_{hole}-T_{dark}}. \qquad (S11)$$

To calculate the amount of absorbed light we also measure reflection

$$R = \frac{R_{sample}-R_{hole}}{R_{mirror}-R_{dark}} \qquad (S12)$$

and use $Abs = 1 - R - T$ to calculate the amount of absorbed light (Fig. S8c). Overall we find very good agreement between this new transmission measurements (blue) and previously obtained data from nanomechanical spectroscopy (red).

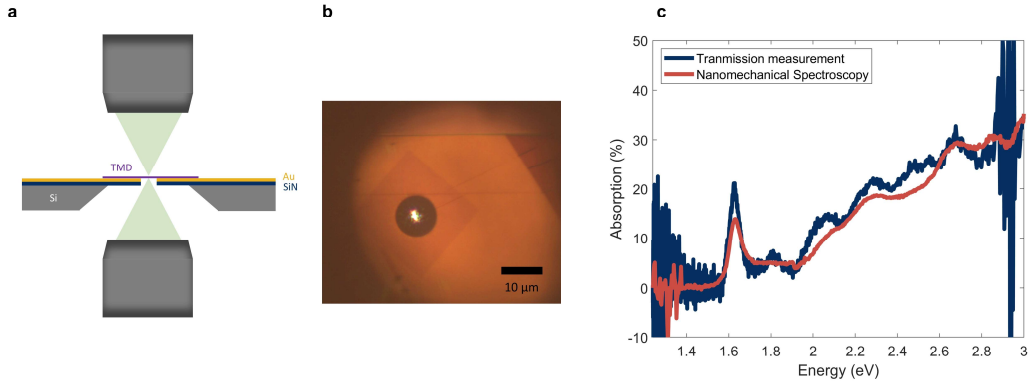

**Figure 8. Optical transmission as Benchmark. a,** Schematic side view of transmission measurement. **b**, Device #1 from the main paper with a beam from a coherent white light source focused on its center. **c** Comparing nanomechanical spectroscopy to optical transmission measurements. We find very good agreement in the determined absorption of the 2D material.

## 8. Reflection measurements

The setup presented in the main text also allows us to perform reflection measurements. We block the reference arm, turn off the probe laser and then and use our tunable excitation light source to sweep the wavelength whilst recording the reflected signal off our sample using a chopper (920Hz) and the lock-in amplifier (Fig. S9a, green). We then subtract spectra from that from an empty hole as shown in Fig. S8a (Fig. S9a, blue) and normalize the data by dividing by a "100% reflection reference", which we obtain measuring reflection of a silver mirror (Fig. S9a, red) with known reflection properties (Thorlabs PF10-03-P01). The resulting reflection data is shown in Fig. S9b.

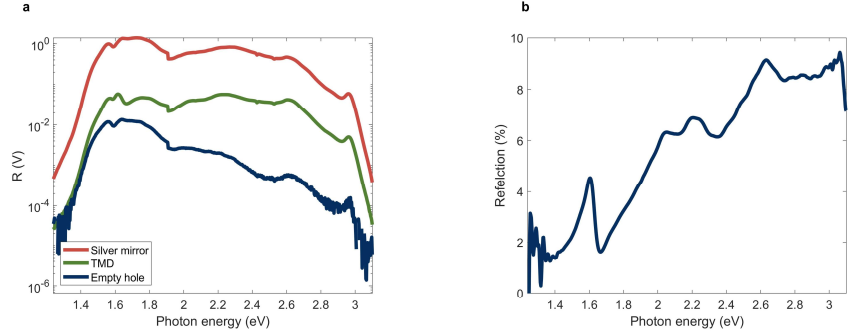

**Figure S9 Optical reflection measurements a)** Reflection of device #1 (3L WSe$_2$), a silver mirror, corresponding to our 100% reflection reference and an empty hole **b)** Resulting reflection data for the TMD material.

## 9. Obtaining the dielectric function

Reflection and transmission of electromagnetic waves was computed with the transfer matrix formalism. Two types of matrices are required: a propagation matrix $P$ and a boundary matrix $T$. The propagation matrix contains elements responsible for phase change inside a material

$$P(\lambda, n, d) = \begin{pmatrix} e^{2\pi i n d/\lambda} & 0 \\ 0 & e^{-2\pi i n d/\lambda} \end{pmatrix}, \quad (S13)$$

where $n$ is complex refractive index of the material, $\lambda$ wavelength of light, $d$ is the thickness of the material. Whereas the boundary matrix depends on the refractive indices on both sides of the boundary $n_1$ and $n_2$:

$$T(n_1, n_2) = \frac{1}{t_{12}} \begin{pmatrix} 1 & r_{12} \\ r_{12} & 1 \end{pmatrix}. \quad (S14)$$

The $t_{12} = \frac{2n_1}{n_1+n_2}$ is a Frensel transmission coefficient for oblique incidence and the $r_{12}$ is a reflection coefficient $\frac{n_1-n_2}{n_1+n_2}$. The overall transfer matrix $M$ of vacuum suspended TMDC yields

$$M = T(n_{vacuum}, n_{TMDC}).P(\lambda, n_{TMDC}, d).T(n_{TMDC}, n_{vacuum}). \quad (S15)$$

For each wavelength, we compute the refractive index $n_{TMDC} = n + Ik$ using matrix elements. The system of two equations is solved for two variables $n, k$.

$$\begin{cases} Trans = 1 - Abs - Refl = 1/|M_{11}|^2 \\ Refl = |M_{21}|^2/|M_{11}|^2 \end{cases}, \quad (S16)$$

$Abs, Refl$ are experimentally obtained absorption and reflection, respectively. The dielectric function $\varepsilon$ is obtained using relation $\varepsilon = n^2$.

### 10. RPA and BSE calculations

To determine the theoretical response function the ground-state of the material was first calculated using density functional theory (DFT). Within DFT, the exchange-correlation energy was approximated by the local density approximation (LDA), which is well known for underestimating the bandgap of insulators and semiconductors. In order to estimate the experimental direct bandgap $G_0W_0$ calculations[25] were performed and the DFT band-structure was then corrected by the scissor operator to obtain the correct direct bandgap.

This corrected band-structure was then used to determine the response function of the material. In order to account for excitonic effects the Bethe Salpeter equation (BSE) was solved[26]. Solving the BSE is computationally very demanding and hence the BSE Hamiltonian was diagonalized in a restricted active space of a few bands around the Fermi level. However, the consequence of this restriction is that the response function is only determined in a limited low energy window around the band-gap. In order to obtain the response function at higher energies, where excitonic effects are negligible, we use the so-called random-phase approximation (RPA) within linear response time-dependent density functional theory (TDDFT).[27,28] This procedure does not account for excitonic effects, but bands up to 100 eV above the Fermi energy are included and is an accurate method for determination of response function away from the band-gap energies.

Computational parameters: Spin-orbit coupling was included for all calculations. For the DFT calculations the in-plane lattice parameter for WSe$_2$ (MoS$_2$) was 3.28 Å (3.16 Å) with an interlayer spacing of 6.48 Å (6.15 Å), a distance of 3.34 Å (3.17 Å) between the chalcogens in each layer, and vacuum spacing between top and bottom layers of at least 12 Å for both the tri- and tetra-layer calculations. A k-point grid of 30x30x1 was used in all cases. The BSE hamiltonian was diagonalized in the restricted active space of 8 valence and 8 conduction states around the Fermi level. In order to account for many-body effects we have performed a single shot, finite temperature (a temperature of 500 K was used), all electron, spin-polarized GW calculations, where the spectral function on the real axis is constructed using a Pade approximation. Spin-orbit coupling was included in the GW calculations and a Matsubara cut-off of 12 Ha was used. All calculations were performed using state-of-the-art, all-electron, full-potential code Elk.[29]

## 11. Determination of sensitivity via Allan deviation:

The Allan deviation is defined as:[31]

$$\sigma_A^2(t) = \frac{1}{2(N-1)f_0^2}\sum_{i=2}^{N}(f_i - f_{i-1})^2 \qquad (S15)$$

where $f_i$ is the average frequency measured over the $i$th time interval of length $t$. We perform time stability measurements (Fig. 5b, main paper) of the resonance frequency with the heating laser turned off using a PLL with a bandwidth BW = 2.5 kHz. We extract $\sigma_A$ and find $\sigma_A < 5 \cdot 10^{-7}$ over a broad range (Fig. 5c, main paper). Plugging $\frac{\Delta f}{\Delta P} = $ 792 Hz/µW, $f_0$ = 4.6702 MHz and an optimal $\sigma_A = 2.426 \cdot 10^{-7}$ at a sampling period of $t$ = 4 ms into equation S16, we calculate $= 90 \frac{pW}{\sqrt{Hz}}$. The measurement fulfils the condition of $t \gg \frac{1}{BW}$.

$$\eta = \frac{\sigma_f \sqrt{t}}{f_0\left(\frac{\Delta f}{f_0 \Delta P}\right)} = \frac{\sigma_A \sqrt{t} f_0}{\frac{\Delta f}{\Delta P}}, \qquad (S16)$$